\documentclass[reprint, prl]{revtex4-1}

\usepackage{scrextend}

\usepackage{amsmath}
\usepackage{amssymb}
\usepackage{graphicx}
\usepackage{dcolumn}
\usepackage{bm}
\usepackage{color}
\usepackage[usenames,dvipsnames]{xcolor}

\usepackage{bm}
\usepackage{graphicx}
\usepackage{ulem}

\def\be{\begin{equation}}
\def\ee{\end{equation}}
\def\ber{\begin{eqnarray}}
\def\eer{\end{eqnarray}}
\def\bwt{\begin{widetext}}
\def\ewt{\end{widetext}}

\begin{document}

\title {Plasmons in dimensionally mismatched Coulomb coupled graphene systems}
\author{S. M. Badalyan}
\affiliation{Center for Nanostructured Graphene (CNG), Department of Micro and Nanotechnology, Technical University of Denmark, DK-2800 Kongens Lyngby, Denmark}
\author{A. A. Shylau}
\affiliation{Center for Nanostructured Graphene (CNG), Department of Micro and Nanotechnology, Technical University of Denmark, DK-2800 Kongens Lyngby, Denmark}
\author{A. P. Jauho}
\affiliation{Center for Nanostructured Graphene (CNG), Department of Micro and Nanotechnology, Technical University of Denmark, DK-2800 Kongens Lyngby, Denmark}

\date{\today}

\begin{abstract}
We calculate the plasmon dispersion relation for Coulomb coupled metallic armchair graphene nanoribbons and doped monolayer graphene. The crossing of the plasmon curves, which occurs for uncoupled 1D and 2D systems, is split by the interlayer Coulomb coupling into a lower and an upper plasmon branch. The upper branch exhibits an unusual behavior with endpoints at finite $q$. Accordingly, the structure factor shows either a single or a double peak behavior, depending on the plasmon wavelength. The new plasmon structure is relevant to recent experiments, its properties can be controlled by varying the system parameters and be used in plasmonic applications.
\end{abstract}

\pacs{71.45.Gm; 73.20.Mf; 73.21.Ac; 81.05.ue}


\maketitle

\paragraph{Introduction}

Collective self oscillations of free electronic charges, known as plasmons ~\cite{Pines1966}, have been of considerable  experimental~\cite{Maier2007} and theoretical~\cite{GV2005} interest for several decades. Plasmon properties depend on the dimensionality of electronic systems \cite{DSarma2009,Zoua2001}. In low dimensions plasmons have been intensively studied in individual electronic systems in semiconductors~\cite{Ando1982} and graphene~\cite{Stauber2014}. 
Recently, an interesting concept has been introduced for studying the plasmonic response of graphene using gratings generated by surface acoustic phonons~\cite{Farhat,Schiefele}.
Rapid developments in graphene plasmonics~\cite{Polini2012,Goncalves2016} hold a great promise of new functionalities of plasmons~\cite{Low2014ACS}, particularly because of their gate-tunability~\cite{Vakil2011,Fang2012,Yan2012}, long lifetime~\cite{Yan2013}, and the extreme confinement of the optical field~\cite{Geim2008,Jablan2009,Koppens2012}.

The dimensionality reduction creates a new class of Coulomb coupled electronic systems~\cite{Macdonald1997} in spatially separated double~\cite{Sarma2009,Stauber2012,Profumo2012,SMB2012} and multilayers~\cite{Sarma1982,SMB2013,Rodin2015}. These structures, where the electronic subsystems may have different dimensionality, open up new ways for identifying the influence of the dimensionality mismatch on interaction phenomena. Thus far, these effects have received only limited attention~\cite{Picciotto2001,Horing2009,Sirenko1992}, but the very recent experiments on Coulomb drag~\cite{Kim2017} may change the situation drastically. (For early related theoretical work see Ref.~\cite{Lyo2003}.)  Hitherto,  the investigations of plasmons in one-dimensional (1D) and two-dimensional (2D) structures have been restricted to electronic multilayers with subsystems of equal dimensionality.

In the present article we develop a theory to describe the dynamical screening in electronic bilayers consisting of Coulomb coupled subsystems of different dimensionality, and use it to identify the structure of the plasmon spectrum in spatially separated metallic armchair graphene nanoribbons and monolayers of graphene.
Due to the dimensionality mismatch, the energy dispersions of plasmons in the individual structures of graphene cross at intermediate energies and momenta.
We find that the interlayer Coulomb coupling drastically changes the plasmon spectrum inducing a new structure in 1D-2D electronic systems. These hybrid bilayers are effectively one-dimensional systems and hence do not support the existence of graphene-like plasmon excitations with a square-root dispersion in the long wavelength limit. Instead, the plasmon spectrum consists of lower and upper split-off branches, which exhibit {\it endpoints} on the dispersion curves.
Therefore, depending on the plasmon wavenumber, the structure factor exhibits either a single-peak or a double-peak behavior as a function of the bosonic frequency. We identify also a narrow window of plasmon momenta where the peak with higher energy can itself be split.
The energy splitting of these hybrid plasmons and their other properties can be controlled by varying the interlayer spacing, the nanoribbon width, and the carrier density in graphene.
Our choice of the metallic armchair nanoribbon as the 1D subsystem is motivated by the simple analytical form of the electron polarization function~\cite{SMB2015}. We emphasize, however, that dispersion properties of 1D plasmons in doped zigzag and semiconducting armchair graphene nanoribbons are essentially the same~\cite{BF2007,Castro2009}. Therefore, our results have a generic validity and are applicable to different types of 1D-2D hybrid systems.
The hybrid 1D-2D plasmons can make an essential contribution to the drag resistance, therefore our results are directly relevant to the recent Coulomb drag experiments where a metallic carbon nanotube is coupled to a monolayer of graphene.~\cite{Kim2017}
We propose that the dimensionality mismatch in electronic multilayers can be useful in designing other plasmon applications.

\paragraph{Theoretical model}

\begin{figure}[t]
\includegraphics[width=.9\columnwidth]{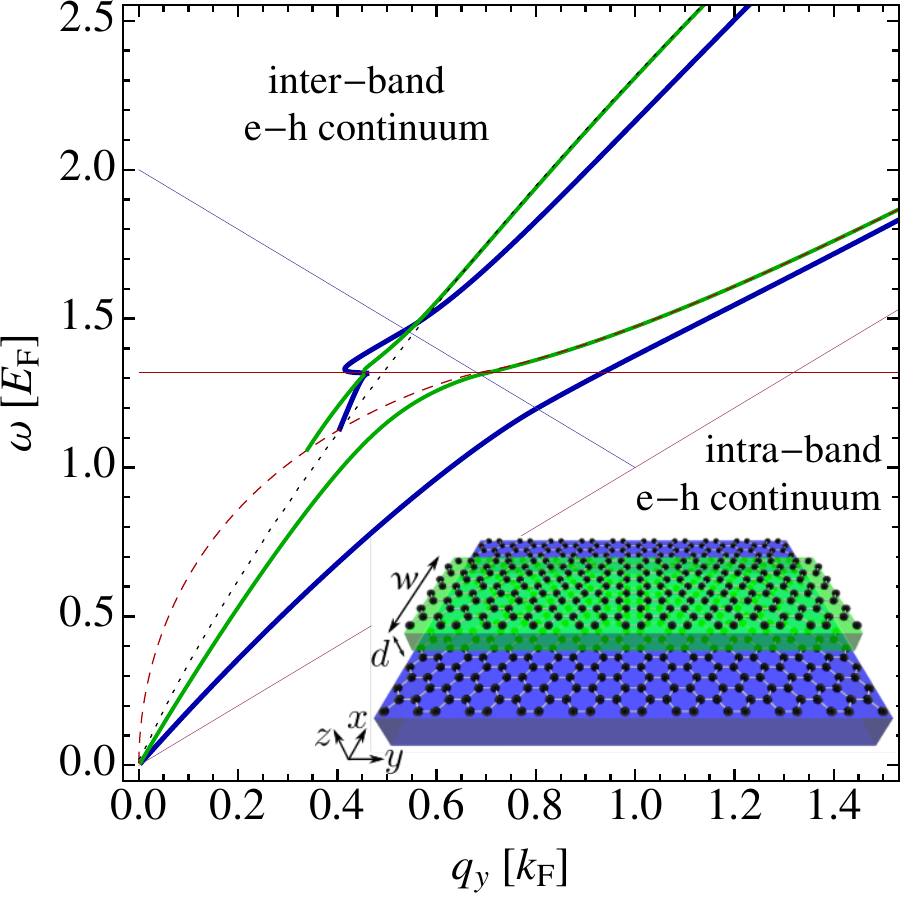}
\caption{
Plasmon dispersion in Coulomb coupled metallic armchair graphene nanoribbons and monolayer graphene. The green ($d=16$ nm) and blue ($d=2$ nm) solid curves show the energy dispersions of the upper and lower branches of the hybrid plasmons for two representative spacings.
The dotted and dashed curves show the uncoupled 1D and 2D plasmon dispersions as a guide to the eye.
The horizontal thin line at $\omega\approx 1.32 E_F$ corresponds to the energy at which the 2D plasmon enters into the 2D interband EHC of graphene. The other two thin lines show the boundaries of the 2D inter and interband EHC in monolayer graphene.
The nanoribbon width is $w=12$ nm and the electron areal density in graphene monolayer $n_\text{gr}=3\times10^{11}$ cm$^{-2}$ Inset: schematics of a dimensionally mismatched graphene nanostructure.
}
\label{fig2}
\end{figure}

The double-layer structure under consideration here consists of a metallic armchair graphene nanoribbon ("layer 1"), with a finite width $w$ in the transverse $x$-direction, and a monolayer of graphene ("layer 2"), which are spatially separated in the $z$-direction with a spacing $d$ ({\it cf.} inset in Fig.~\ref{fig2}).
The system is nonuniform along the $x$ and $z$ directions and we find the plasmon excitations from the poles of the Fourier transform of the exact Coulomb propagator in the translationally invariant $y$-direction as a function of the momentum $q_y$ and the energy $\omega$.
In real space ${\bf r}=(x,y,z)$ the exact Coulomb propagator, $W_{ij}({\bf r_1, r_2}|\omega)$, satisfies the integral Dyson equation
\begin{eqnarray}\label{DERS}
&&\hat{W}({\bf r_1, r_2}|\omega)=\hat{V}({\bf r_1, r_2})
\\ &&+
\int \hat{V}\left({\bf r_1,\bar{r}_1}\right) \hat{\Pi}\left({\bf \bar{r}_1,\bar{r}_2}|\omega\right)\hat{W}\left({\bf \bar{r}_2,r_2}|\omega\right)d\bar{\bf r}_1 d\bar{\bf r}_2~. \nonumber
\end{eqnarray}
Here, the propagator $\hat{W}({\bf r_1, r_2}|\omega)$, the electron polarization function, $\hat{\Pi}\left({\bf \bar{r}_1,\bar{r}_2}|\omega\right)$, and the bare Coulomb interaction, $\hat{V}\left({\bf r_1,r_2} \right)$,  
are $2\times2$ matrices with respect to the layer indices $i, j=1,2$. The layers are assumed sufficiently far apart so that the interlayer tunneling and the nondiagonal elements of $\hat{\Pi}\left({\bf \bar{r}_1,\bar{r}_2}|\omega\right)$ are negligibly small. Within the random phase approximation the diagonal elements in the $i_{\text{th}}$ layer are
\begin{eqnarray}\label{PFRS}
\Pi_{ii}\left({\bf r}_1,\left. {\bf r}_2\right|\omega \right)&=&\sum_{\mu,\nu}{\Psi _{\nu   }^i}^*\left({\bf r}_2\right)\Psi _{\nu   }^i\left({\bf r}_1\right) \Psi _{\mu }^i\left({\bf r}_2\right) {\Psi _{\mu   }^i}^*\left({\bf r}_1\right) \nonumber
\\ &\times&
\frac{f\left(E_{\mu   }^i\right)-f\left(E_{\nu }^i\right)} {E_{\mu }^i -  E_{\nu }^i+\omega + i 0 }~.
\end{eqnarray}
and can be calculated using the electron wave functions, $\Psi_{\mu}\left({\bf r}\right)$, and the 
energy spectra, $E_{\mu}$, in graphene nanoribbons and monolayers of graphene~\cite{Castro2009}. Here $f\left(E_{\mu }\right)$ is the Fermi function.
The indices $\mu,\nu$ are combined quantum numbers, which describe the electron motion in the respective layer.

In armchair graphene nanoribbons $\mu=(n,s,k_y)$ where the transverse quantization subband index is an integer, $n=0,\pm 1,\dots$, and the chirality index $s=\pm1$.
The conserved momentum $k_{y}$ corresponds to the translational invariant $y$ direction.
We assume that the Fermi energy and temperature are smaller than the transverse quantization energy, $E_{\text{F}}, T\ll \pi v_{\text{gr}}/ w$. 
Here units $k_B=\hbar=1$ are used, and $v_{\text{gr}}$ the velocity of graphene. For  carrier densities in nanoribbons, corresponding to the areal density $n_\text{gr}=3\times10^{11}$ cm$^{-2}$, and for $w=12$ nm, we have $\pi v_{\text{gr}}/w\approx 1996$ K and $E_{\text{F}}=\pi n_\text{gr} w v_\text{gr}/4\approx 215.6$ K.
%
%
In this regime experimentally relevant structures are metallic armchair graphene nanoribbons with the single-particle energy spectrum, $E_{s}(k_{y}) = s v_{\text{gr}} k_{y}$, of 1D Dirac fermions.

In monolayer graphene the quantum number $\mu=(s,{\bf p})$ describes the 2D electron spinor states in the $(x,y)$ plane with the in-plane momentum ${\bf p}$ and the single-particle Dirac  spectrum $E_{s}\left( p \right) = s v_{\text{gr}} p$.
We neglect electronic transitions due to intervalley scattering (for large values of the transferred momentum interlayer Coulomb interaction is small) and take into account the valley index via the degeneracy factor in the definition of the Fermi momentum and energy.

\paragraph{Solution of the Dyson equation}

We rewrite the Dyson equation (\ref{DERS}) for the Fourier components of the exact interactions in the $y$-direction, $\bar{W}_{ij}\left(x_1,x_2 | q_y |\omega \right)$, which are also weighted by the carrier densities in the $z$ direction, to take into account the carrier localization in the respective layers.
Next we average $\bar{W}_{ij}\left(x_1,x_2 | q_y |\omega \right)$ over the transverse coordinates of electrons, and introduce the notation $\tilde{W}_{11}\left(q_y | \omega \right)=\int^{w/2}_{-w/2}\int^{w/2}_{-w/2} \bar{W}_{11}\left(x_1,x_2 | q_y |\omega \right) dx_1dx_2/w^2$ and $\tilde{W}_{21}\left(q_x, q_y | \omega \right)=\int^{\infty}_{-\infty} \tilde{W}_{21}\left(x_1 | q_y |\omega \right) e^{i q_x x_1} dx_1$ with $\tilde{W}_{21}\left(x_1 | q_y |\omega \right)=\int^{w/2}_{-w/2} \bar{W}_{21}\left(x_1,x_2 | q_y |\omega \right) dx_2/w$.
Note that in contrast to the bare Coulomb interaction $\bar{V}_{ij}\left(x_1-x_2 | q_y\right)$, the exact interactions $\bar{W}_{ij}\left(x_1,x_2 | q_y |\omega \right)$ depend on the coordinates $x_1,x_2$ separately.
Then, the system of equations for the components $\tilde{W}_{11}\left(q_y | \omega \right)$ and $\tilde{W}_{21}\left(q_x, q_y | \omega \right)$ is represented as
\begin{widetext}
\begin{subequations}
\begin{eqnarray} \label{SYS}
&&\tilde{W}_{11}\left(q_y | \omega \right)=\tilde{V}^{1\text{D}}_{11}\left(q_y\right)
+\tilde{V}_{11}^{1\text{D}}\left(q_y\right) \Pi _{11}^{1\text{D}}\left(q_y, \omega\right)  \tilde{W}_{11}\left(q_y|\omega \right)
+\frac{1}{L} \sum_{q_x} I\left(q_x\right) V_{12}^{2 \text{D}}\left( q\right) \Pi_{22}^{2 \text{D}}\left(q, \omega \right) \tilde{W}_{21}\left(q_{x},q_{y} | \omega \right)
\\
&&\tilde{W}_{21}\left(q_{x},q_{y} | \omega \right)=I \left(q_x\right) V^{2\text{D}}_{21}\left(q\right)
+ I \left(q_x\right) V_{21}^{2\text{D}}\left(q\right)  \Pi _{11}^{1\text{D}}\left(q_y, \omega\right)  \tilde{W}_{11}\left(q_y | \omega \right)
+V_{22}^{2 \text{D}}\left( q\right) \Pi_{22}^{2 \text{D}}\left(q, \omega \right) \tilde{W}_{21}\left(q_x, q_y | \omega \right)
\end{eqnarray}
\end{subequations}
\end{widetext}
with the form factor $I\left(q_x\right)=2\sin(q_x w/2)/q_x w$. We introduce also the averaged bare interaction in graphene nanoribbons as $\tilde{V}^{1 \text{D}}_{11} \left(q_y\right)=\int^{w/2}_{-w/2}\int^{w/2}_{-w/2}dx_1dx_2 V_{11} \left(x_1-x_2 | q_y \right)$ where $V_{11} \left(x_1-x_2 | q_y \right)=2 e^2/\epsilon_\text{eff} K_0 \left(q_y \left| x_1-x_2 \right|  \right)$ with $\epsilon_\text{eff}$ the effective low frequency dielectric function of the background dielectric medium, and $K_0 $ is the modified Bessel function of the second kind.
The functions $V^{2 \text{D}}_{ij}\left(q\right)=2\pi e^2 e^{-|i-j| q d} / \epsilon_\text{eff} q $ are the 2D Fourier transforms of the bare intra and interlayer Coulomb interaction with $q=\sqrt{q_x^2+q_y^2}$.

Then, we find the solution of the Dyson equation as
%
\begin{eqnarray}\label{SOL}
\tilde{W}_{11}\left(q_y | \omega \right)&=&\frac{\tilde{V}^{1\text{D}}_{\text{eff}}\left(q_y | \omega\right)}{\varepsilon_{\text{1D-2D}}\left(q_y,\omega\right)},\nonumber\\
\tilde{W}_{21}\left(x | q_y | \omega\right) &=&\frac{\tilde{V}^{1\text{D}-2\text{D}}_{\text{eff}}\left(x|q_y| \omega\right)}{\varepsilon_{\text{1D-2D}}\left(q_y,\omega\right) }~.
\end{eqnarray}
Here the central quantity is the dynamical screening function of the hybrid 1D-2D electronic system
\begin{eqnarray}\label{EPS}
\varepsilon_{\text{1D-2D}}\left(q_y,\omega \right)&=& \varepsilon_\text{1D}\left(q_y, \omega \right)\nonumber\\
&\quad&-
{\cal Q}_{\text{1D-2D}}\left(q_y,\omega \right) \Pi_{11}^{1 \text{D}}\left(q_y, \omega \right)
\end{eqnarray}
where 
\begin{eqnarray}\label{Q}
{\cal Q}_{\text{1D-2D}}\left(q_y,\omega \right)=\frac{1}{L} \sum _{q_x} \frac{I^2 \left(q_x\right) V^{2\text{D}}_{12}\left(q\right)^{2}\Pi_{22}^{2 \text{D}}\left(q, \omega \right)}{\varepsilon_{2 \text{D}}\left(q, \omega \right)} ~.
\end{eqnarray}
The 1D and 2D interlayer dynamical screening functions (the Lindhard polarization functions) in graphene nanoribbons~\cite{BF2007,SMB2015,Castro2009} and monolayers of graphene~\cite{Guinea2006,Sarma2007,Pyatkovskiy,Kotov2012} are, respectively, $\varepsilon_\text{1D}\left(q_y, \omega \right) =1 - \tilde{V}^{1 \text{D}}_{11}\left(q_y \right)\Pi^{1 \text{D}}_{11}\left(q_y, \omega \right)$
and $\varepsilon_\text{2D}\left(q, \omega \right) =1 - V^{2 \text{D}}_{22} \left(q\right) \Pi^{2\text{D}}_{22}\left(q, \omega \right)$.
In Eq.~(\ref{SOL}) we define the effective intraribbon and interlayer interactions, respectively, as $\tilde{V}^{1 \text{D}}_\text{eff}\left(q_y | \omega \right)=\tilde{V}^{1 \text{D}}_{11}\left(q_y \right)+{\cal Q}_{\text{1D-2D}}\left(q_y,\omega \right)$ and $\tilde{V}^{1\text{D}-2\text{D}}_{\text{eff}}\left(x|q_y| \omega\right)=1/L \sum_{q_x} e^{-i q_x x} I\left(q_x\right) V^{2\text{D}}_{12}\left(q\right)/\varepsilon_{2\text{D}}\left(q,\omega\right)$.
Notice that the effective interactions have no poles as a function of $q_y$. 
In the limit of vanishing interlayer interaction $V^{2\text{D}}_{12}\left(q\right)\rightarrow0$ for large values of $d$ and/or $x$, the electronic subsystems in nanoribbon and monolayer graphene become independent. In addition to the 1D momentum $q_y$, the 2D momentum $q$ becomes a well-defined conserved quantum number in monolayer graphene because of the recovered 2D translational invariance. Then, the full screening function is represented as a simple product of its 1D and 2D parts. The poles of the 2D Fourier transformed propagator $\tilde{W}_{21}\left(q_{x},q_{y} | \omega \right)$ as a function of $q$ are given by $\Re\varepsilon_{2\text{D}}\left(q,\omega\right)=0$ and determine the square-root spectrum of plasmons in an individual graphene sheet. In this limit, Eq.~(\ref{EPS}) is reduced to the 1D screening function of an individual graphene nanoribbon and determines the dispersion of 1D plasmons as a function of $q_y$.  With a decrease of $x$ and $d$ the 1D-2D coupling is recovered and the hybrid 1D-2D modes govern the plasmon spectrum as a function of $q_y$. Further our discussion is restricted only to these new hybrid plasmon modes.

Thus, Eqs.~(\ref{SOL})-(\ref{Q}) allow us to describe the dynamical screening phenomena and to obtain the plasmon structure in Coulomb coupled electronic bilayers, consisting of subsystems with different dimensionality. These formulae are general and allow us to describe hybrid structures with a different type of graphene and conventional electronic subsystems as well as mixed structures. Microscopic details of the subsystems determine the functional forms of the 1D and 2D polarization functions and the form factor $I\left(q_x\right)$.~\footnote{Using the 1D and 2D polarization functions and the form factor $I\left(q_x\right)$ for conventional electron gases, Eqs.~(\ref{SOL})-(\ref{Q}) reproduce in the static limit the results of Ref.~\cite{Lyo2003}, obtained using the diagrammatic technique.
}

\paragraph{Plasmon dispersions}
Dispersive properties of hybrid plasmons in 1D-2D electronic bilayers are determined by the zeroes of the real part of the dynamical screening function~\footnote{Eq.~(\ref{DE}) determines the poles of all the four components of the exact propagator $\tilde{W}_{ij}$.}
\begin{equation}\label{DE}
\Re~\varepsilon_{\text{1D-2D}}\left(q_y,\omega \right)=0~.
\end{equation}
To obtain a solution to this equation we note that the summand in Eq.~(\ref{Q}) has different analytical properties in the high-energy, $\omega>\omega_{2\text{D}}(q_y)$, and the low-energy, $\omega<\omega_{2\text{D}}(q_y)$, regions where $\omega_{2\text{D}}(q_y)$ is the 2D plasmon energy in graphene with $q_x=0$ ({\it cf.} the dashed curve in Fig.~\ref{fig2}). In the $\omega>\omega_{2\text{D}}(q_y)$ region 
$\varepsilon_{2 \text{D}}\left(q, \omega \right)$ always has a zero as a function of $q_x$, corresponding to the plasmon energy in an uncoupled graphene sheet. Therefore, in this regime we calculate $Q_{\text{1D-2D}}\left(q_y, \omega\right)$ taking its principal value numerically.

In the low-energy regime, $\omega<\omega_{2\text{D}}(q_y)$, the integrand in Eq.~(\ref{Q}) is not singular and one can calculate the energy dispersion of the lower branch of the hybrid plasmon numerically from Eqs.~(\ref{EPS})-(\ref{DE}). Analytically, in the long wavelength limit $q_y w\ll 1$ and $q_y d\ll 1$, we search for the lower plasmon branch in the part of the spectrum close to the  plasmon energy in individual graphene nanoribbons, $\omega\approx \omega_{1\text{D}}\left(q_y\right)$. As far as $\omega_{1\text{D}}\left(q_y\right)\ll \omega_{2\text{D}}\left(q_y\right)<\omega_{2\text{D}}\left(q \right)$, we use the static  approximation for $\Pi_{2 \text{D}}\left(q, \omega \right)\approx \Pi_{2 \text{D}}\left(q, 0 \right)$. 
%
%
Assuming that $d\ll w$, we find the energy dispersion of the lower plasmon in the long wavelength limit as
\begin{eqnarray}\label{out}
\omega_-\left(q_y\right)\approx 2  \left[ 2\alpha_\text{gr} \frac{d}{w} \right]^{1/2} v_\text{gr} q_y
\end{eqnarray}
while in structures with $d\gg w$, it is given by
\begin{eqnarray}
\omega_-\left(q_y\right)\approx 2  \left[ \frac{\alpha_\text{gr}}{\pi} \ln\left({2\frac{d}{w}}\right) \right]^{1/2} v_\text{gr} q_y~.
\end{eqnarray}
The energy of the lower hybrid plasmon is thus linear in $q_y$ and its velocity in the limit of $d\ll w$ is larger by a factor of $\sqrt{2}$ than the velocity of the out-of-phase plasmon in double graphene nanoribbons in the same limit~\cite{SMB2015}.

\begin{figure}[t]
\includegraphics[width=.49\columnwidth]{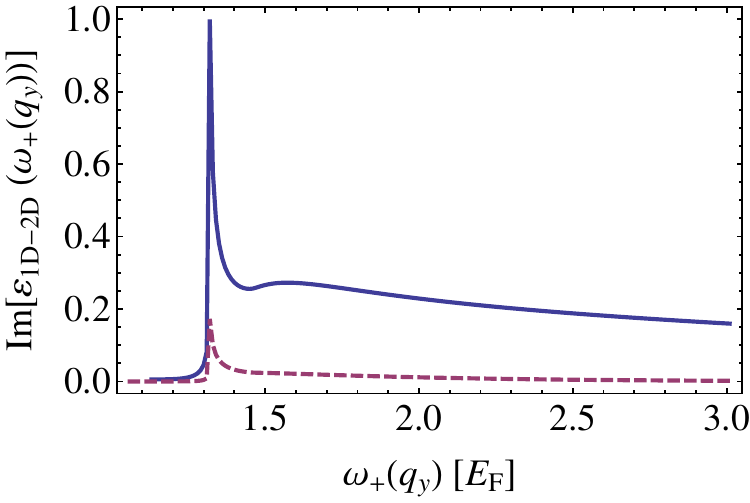}
\includegraphics[width=.49\columnwidth]{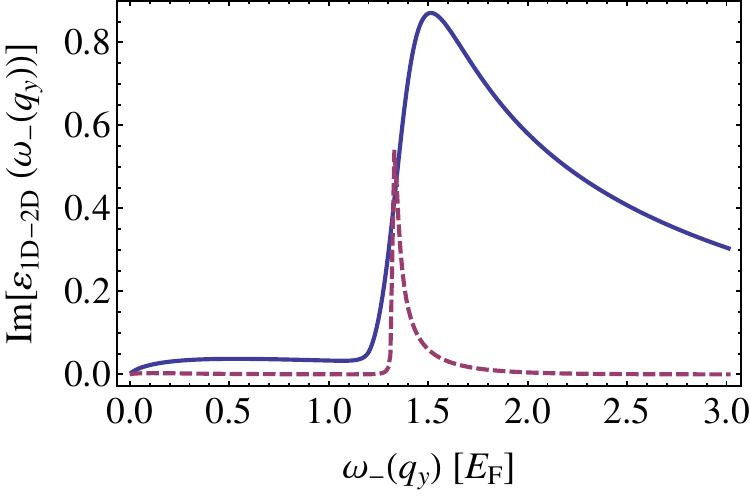}
\caption
{
The imaginary part of the screening function in 1D-2D electronic systems along the upper (left) and the lower (right) hybrid plasmon branches. The solid and dashed curves correspond to interlayer spacing of $d=2$ and $16$ nm.  The values of other parameters are the same as in Fig.~\ref{fig2}.
}
\label{fig3}
\end{figure}

\begin{figure*}[t]
\includegraphics[width=.49\columnwidth]{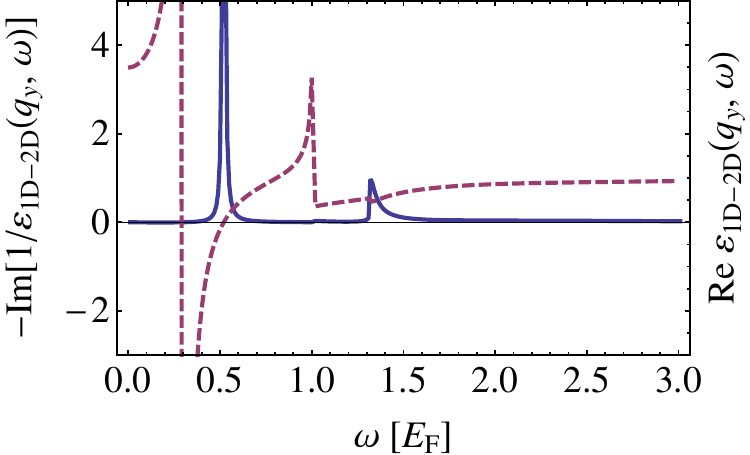}
\includegraphics[width=.49\columnwidth]{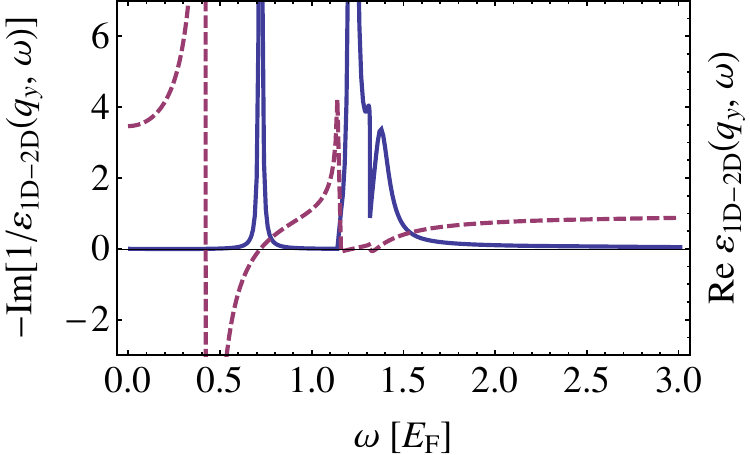}
\includegraphics[width=.49\columnwidth]{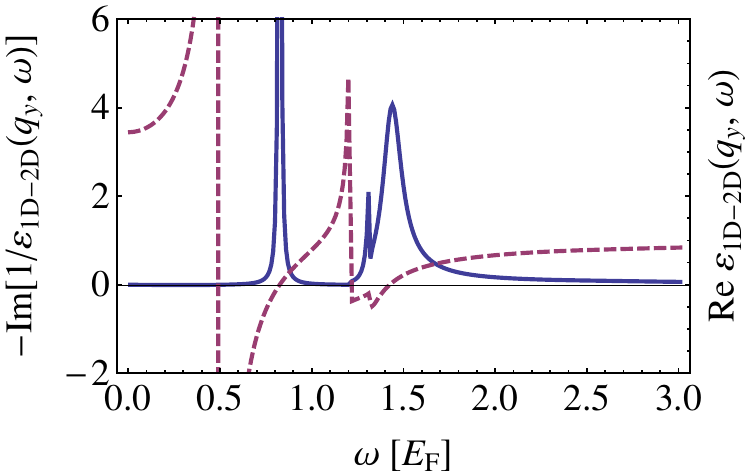}
\includegraphics[width=.49\columnwidth]{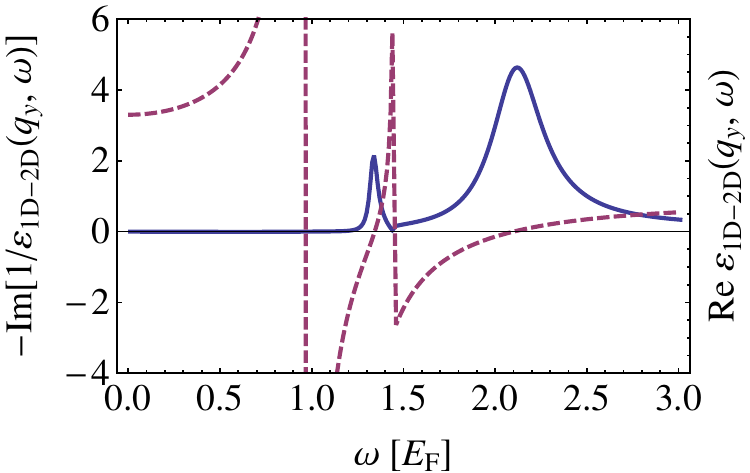}
\caption
{
The dynamical structure factor (solid curve) and the real part of the screening function (dashed curve) in 1D-2D electronic bilayers as a function of the bosonic frequency $\omega$. In the panels from left to right the plots are calculated, respectively, for bosonic momenta $q_y\approx 0.3, 0.43, 0.51$, and $ 0.96 k_F$ and for interlayer spacing $d=2$ nm. The values of other parameters are the same as in Fig.~\ref{fig2}.
}
\label{fig4}
\end{figure*}

\paragraph{Numerical calculations}

In Fig.~\ref{fig2} we plot the plasmon spectrum in the hybrid 1D-2D electronic system, which we calculate assuming that 
$\varepsilon_{2 \text{D}}\left(q, \omega\right)=\Re\varepsilon_{2 \text{D}}\left(q, \omega\right)$. In Eqs.~(\ref{EPS})-(\ref{Q}) we use the explicit $T=0$ expressions of the 1D and 2D polarization functions, respectively, from Refs.~\cite{SMB2015} and \cite{Sarma2007}.
As seen in the figure, the bare plasmon dispersions of the 1D (dotted line) and 2D (dashed line) cross around $q_y=0.4 k_F$. The interlayer Coulomb coupling splits this crossing and induces a new plasmon structure of the upper and lower plasmon modes in the coupled 1D-2D system. This hybrid system is effectively one-dimensional: the plasmon spectrum does not support the $\sqrt{q}$ mode, characteristic for 2D, in the long wavelength limit. The upper branch of the hybrid plasmon has an endpoint of the dispersion curve whose position varies with the system parameters within the intermediate values of $q_y$ and $\omega$, but remains on the boundary $\omega=\omega_{2\text{D}}(q_y)$ separating the high-energy and the low-energy regions. The upper plasmon shows also a singular behavior at the critical energy $\omega_\text{c}$, at which the individual 2D plasmon enters into the 2D electron-hole continuum (EHC). This is due to the singularity of the 2D dielectric function (the second derivative of $\varepsilon_{2 \text{D}}\left(\omega_{2\text{D}}(q)\right)$ has a gap at $\omega_{2\text{D}}(q)=\omega_\text{c}$), an intrinsic feature of the plasmon dispersion in 2D graphene monolayers.
Note that the dispersion curve of the upper plasmon crosses that of the 1D plasmon, i.e. the interlayer interaction effectively vanishes at certain values of $\omega$ and $q_y$. A similar situation takes place in an individual graphene sheet because the polarizability $\Pi^{2\text{D}}_{22}\left(q, \omega \right)$ vanishes at certain intermediate values of $\omega$ and $q$~\cite{Kotov2012}.

Using the full complex 2D dielectric function $\varepsilon_{2 \text{D}}\left(q, \omega\right)=\Re\varepsilon_{2 \text{D}}\left(q, \omega\right)+ i \Im\varepsilon_{2 \text{D}}\left(q, \omega\right)$ modifies markedly the plasmon spectrum only in the region of $\omega>\omega_\text{c}$. The lower plasmon branch acquires an endpoint on the dispersion curve, lying on the boundary $\omega=\omega_{2\text{D}}(q_y)$ at $\omega\approx 1.43 E_\text{F}$ and $\omega\approx 1.32 E_\text{F}$, respectively, for $d=2$ nm and $16$ nm. The energy of the upper plasmon branch becomes about $10 \%$  lower at large values of $q_y$ for $d=2$ nm while for $d=16$ nm, the changes are almost invisible.

In the large $q_y$ limit, the upper plasmon branch behaves as a nanoribbon-like mode, whose energy tends to the energy of the 1D plasmon in uncoupled graphene nanoribbons with an increasing interlayer spacing $d$. The energy of the lower branch of the hybrid plasmon shows a similar trend, but in the opposite low $q_y$ limit.

The interlayer Coulomb coupling modifies strongly also the dissipative properties of the 1D-2D plasmons.
Notice hybrid plasmons propagate in the translationally invariant $y$-direction and the broadening of plasmon peaks describes the plasmon damping in this direction. 
In Fig.~\ref{fig3} we plot the imaginary part of the dynamical screening function, $\Im\varepsilon_{\text{1D-2D}}\left(\omega_{\pm}(q_y) \right)$, for the upper and lower modes, which are calculated using the complex
$\varepsilon_{2 \text{D}}\left(q, \omega\right)$.
In contrast with the behavior of the individual 1D and 2D plasmons, both the upper and lower hybrid plasmon modes are Landau damped in the whole $(\omega,q_y)$-plane of the spectrum. However for both modes, $\Im\varepsilon_{\text{1D-2D}}\left(\omega_{\pm}(q_y) \right)$ is sufficiently small for energies $\omega<\omega_\text{c}$ so that the hybrid plasmons are well defined excitations in this region.
For the upper branch, $\Im\varepsilon_{\text{1D-2D}}\left(\omega_{+}(q_y) \right)$ is rather large outside the 2D EHC of monolayer graphene in the energy region of $\omega>\omega_\text{c}$. Meanwhile, for the lower branch $\Im\varepsilon_{\text{1D-2D}}\left(\omega_{-}(q_y) \right)$ is rather small inside the 2D EHC, but for energies $\omega<\omega_\text{c}$.
For both modes the imaginary part shows peaks at energies immediately above $\omega_\text{c}$ and decreases with an increasing $\omega$, reflecting the behavior of $-\Im\varepsilon^{-1}_{\text{2D}}\left(q,\omega \right)$ in the 2D graphene sheet. As seen, $\Im\varepsilon_{\text{1D-2D}}\left(\omega_{+}(q_y) \right)$ is essentially smaller in structures with larger $d=16$ nm spacing so that in the limit of $q_y d\rightarrow\infty$ the spectrum of the hybrid plasmon recovers the {\it undamped} 1D plasmon in metallic armchair graphene nanoribbons.

The dispersive and dissipative features of the 1D-2D plasmons discussed above determine the behavior of the dynamical structure factor, $S_\text{1D-2D}\left(q_y, \omega \right)=-\Im \varepsilon_\text{1D-2D}^{-1} \left(q_y, \omega \right)$. In Fig.~\ref{fig4} we plot $S_\text{1D-2D}\left(q_y, \omega \right)$ as a function of $\omega$ together with 
$\Re \varepsilon_\text{1D-2D} \left(q_y, \omega \right)$ for four typical values of the momentum $q_y$.
It is seen that for $q_y= 0.3$ the structure factor shows a single peak at $\omega \approx 0.52 E_\text{F}$, corresponding to the lower plasmon and a small feature at $\omega \simeq \omega_c$ that reflects the peaked behavior of $-\Im \varepsilon_\text{2D}^{-1} \left(q, \omega \right)$. For $q_y\approx 0.43 k_\text{F}$, in addition to the lower plasmon peak, the structure factor exhibits two more peaks, corresponding to the upper plasmon branch immediately below and above the critical energy $\omega_c$. The latter is strongly damped and suppressed so the peak structure is asymmetric. Note that there is only a small window of momenta around $q_y\approx 0.43 k_\text{F}$ where the upper branch of the 1D-2D plasmon exhibits a double-peak structure. From the comparison of the structure factor behavior for $q_y\approx 0.51 k_\text{F}$ and $q_y\approx 0.96 k_\text{F}$, we see that the lower plasmon peak becomes suppressed at larger momenta while the upper plasmon peak becomes stronger but broader in structures with $d=2$ nm.

In conclusion, we have developed a theory that describes the dynamical screening in electronic bilayers with a dimensionality mismatch. A new plasmon structure has been found in the hybrid 1D-2D systems of graphene nanoribbons and monolayers of graphene, whose properties can be controlled by varying the interlayer spacing, the nanoribbon width, and the carrier density. The results indicate the potential of hybrid graphene multilayers with a dimensionality mismatch for plasmonic applications.

The Center for Nanostructured Graphene (CNG) is sponsored by the Danish National Research Foundation, Project No. DNRF103.

\end{document}